\documentclass[epj]{webofc}
\usepackage[utf8]{inputenc}
\usepackage[varg]{txfonts}   
\usepackage{booktabs}
\usepackage{xcolor}
\definecolor{darkred}{rgb}{0.4,0.0,0.0}
\definecolor{darkgreen}{rgb}{0.0,0.4,0.0}
\definecolor{darkblue}{rgb}{0.0,0.0,0.4}
\usepackage[bookmarks,linktocpage,colorlinks,
    linkcolor = darkred,
    urlcolor  = darkblue,
    citecolor = darkgreen]{hyperref}
\usepackage[normalem]{ulem}

    \newcommand{\pmn}[1]{\textcolor{green}{#1}}

\usepackage{subfigure}
\wocname{EPJ Web of Conferences}
\woctitle{Lattice2017}
\begin{document}
%
\selectlanguage{english}
\title{%
Charmonia in moving frames
 }
\author{%
\firstname{S.} \lastname{Prelovsek}\inst{1,2,3}\fnsep\thanks{ Speaker, \email{sasa.prelovsek@ijs.si}} \and
\firstname{G.} \lastname{Bali}\inst{1} \and
\firstname{S.}  \lastname{Collins}\inst{1}  \and \firstname{D.} \lastname{Mohler}\inst{4} \and   \firstname{M.} \lastname{Padmanath}\inst{1}\and \firstname{S.} \lastname{Piemonte}\inst{1}\and \firstname{S.} \lastname{Weishaeupl}\inst{1}
}
\institute{%
Institute f\"ur Theoretische Physik, Universit\"at Regensburg, D-93040, Germany  
\and 
Faculty of Mathematics and Physics, University of Ljubljana, 1000 Ljubljana, Slovenia
\and 
Jozef Stefan Institute, 1000 Ljubljana, Slovenia
\and
Helmholtz-Institut Mainz, 55099 Mainz, Germany
\and
Johannes Gutenberg Universit\"at Mainz, 55099 Mainz, Germany
}
\abstract{%
Lattice simulation of charmonium resonances with non-zero momentum provides additional information on the two-meson scattering matrices. However, the reduced rotational symmetry in a moving frame renders a number of states with different $J^P$ in the same lattice irreducible representation. The identification of $J^P$ for these states is particularly important, since quarkonium spectra contain a number of states with different $J^P$ in a relatively narrow energy region.  Preliminary results concerning spin-identification are presented in relation to our study of charmonium resonances in flight on the $N_f=2+1$ CLS ensembles. }
\maketitle
\section{Introduction}\label{intro}

Experiments discovered a number of interesting and exotic hadrons in the charmonium sector. Most of those appear near or above open flavor threshold. They have to be inferred from the two-hadron scattering matrices. So far, only one lattice simulation considered the  charmonium resonances above open charm threshold,  performing an  exploratory extraction of their masses as well as decay widths  \cite{Lang:2015sba}. This study  considered only the total momentum  zero, which  provides only a rather limited  
information on the scattering matrix.  Simulation of the system with non-zero total  momentum (i.e. moving frames) provides additional information on the relevant two-meson scattering matrices. 
This presents the main motivation to consider charmonium(like) states in the moving frames.  

On the other hand, simulations   in the moving frames pose additional challenges  compared to  the rest frame. This holds particularly  for the  charmonium sector, since there are a number of states with different $J^P$ in a narrow energy region.  The main focus of this paper is to address  part of those challenges. 

Here we consider the single-hadron approach and employ only quark-antiquark operators $O=\bar cc$.   The motivation is to identify $J^P$ of “single-hadron” eigenstates.  This will be helpful to spin-identify the narrow states once this study is extended to two-hadron scattering; such a study  is pursued in parallel on the same CLS ensembles. The intermediate results at zero momentum were described at this conference in \cite{PiemonteWeishaeupl:2017lat}.  

\section{The challenge in moving frames}

The charmonium spectrum is dense, as shown for example in Figure 3 of  \cite{Cheung:2016bym}, which is obtained with the single-hadron approach and $O=\bar cc$. There is a number of states with different $J^{PC}$ in a narrow energy region, for example  in the interesting region around $4~$GeV. The purpose is to  identify $J^P$ of observed eigenstates in the moving frames. 

The  reduced rotational symmetry in the moving frame renders a number of states with different $J^P$ in the same irreducible representation, as detailed in the next Section. Parity is  not a good quantum number in moving frames\pmn{,} since inversion transforms a system with momentum $\vec P$  to a system with momentum $-\vec P$. Helicity $\lambda=\vec J\cdot \vec P/ |\vec P|$ is a good quantum number instead. Fortunately, charge conjugation is also a good quantum number  in moving frames, which can be used as an important handle to distinguish states of both $C$-parities. 

The strategy that was often used for identifying $J^P$ of eigenstates in the past was based on the near-degeneracy of energies from different irreducible representations. Let us provide  a  familiar example concerning  $\vec P=\vec 0$ with symmetry group $O_h$: the irreducible representation $\Lambda=T_1^{-}$ contains $J^P=1^{-}$ as well as $3^{-}$ states as shown in Table \ref{tab:irreps}; the $J^{PC}=3^{--}$ state was commonly identified as an  eigenstate that has nearly equal energy in irreducible representations $T_1^{-}$ and  $T_2^{-}$. This procedure to identify $3^{--}$ state   is not very reliable for the  dense charmonium spectrum, since the  $3^{--}$ state has almost degenerate mass as $1^{--}$ state  $\psi(3770)$ (see Figure 3 of  \cite{Cheung:2016bym}). This problem becomes even more severe for moving frames, where even more states with different $J^P$ fall into the same irrep. Therefore, the near-degeneracy of eigen-energies between different irreps is not a reliable criteria  for spin-identificaton in general. 

The framework for reliable  identification of $J^P$ for eigenstates at zero momentum was proposed by Hadron Spectrum Collaboration in  \cite{Dudek:2009qf} and was extended to the systems in flight in  \cite{Thomas:2011rh}.  
To identify $J^P$ of charmonium eigenstates in flight we therefore follow   \cite{Thomas:2011rh}. There it was employed in the study of light iso-vector mesons.  
    
  \begin{table}[t]
\begin{center}
\begin{tabular}{|c| c|ccccc|}
\hline
 $\vec P=\vec 0$ &   $J^P$ & $0^\pm$ & &$1^\pm$ & $2^\pm$ & $3^\pm$\\
$O_h$  & $\Lambda^P$ & $A_1^\pm$ && $T_1^\pm$ & $T_2^\pm,E^\pm$ & $T_{1,2}^\pm,A_2^\pm$ \\
 \hline
 &&&&&& \\
 \hline
 $\vec P\propto \vec e_z$  &  $|\lambda|^{\tilde \eta}$ & $0^+$ & $0^-$ & 1& 2 & 3 \\
$\mathrm{Dic}_4$ & $\Lambda$ & $A_1$ & $A_2$ & $E_2$  & $B_{1,2}$ & $E_2$ \\
 \hline
\end{tabular}
\end{center}
\caption{Upper part lists irreps $\Lambda^P$ of the octaheral group ($\vec P=\vec 0$)  and corresponding  continuum quantum numbers  $J^P$ of states that   they contain.  Lower part concerns momentum $\vec P\propto \vec e_z$ with the symmetry group $\mathrm{Dic}_4$: it lists irreps $\Lambda$ and corresponding values of helicity $|\lambda|$ and   $\tilde \eta=P(-1)^J$ for the states that   they contain (with $P$ in $\tilde \eta$ referring to the rest frame).     \label{tab:irreps}}
 \end{table}

  \section{Helicity states and irreducible representations for $\vec P\propto \vec e_z$}\label{sec:irreps}
  
  Let's consider a system with good quantum numbers $J^P$ in its rest frame, but boosted to a momentum $\vec P$. The good quantum number is helicity $\lambda=\vec J\cdot \vec P/ |\vec P|$ that gives the projection of the spin $\vec J$ to the direction of momentum. For given $J$ it can have the values $\lambda=-J,-J+1,..,J$.  The other quantum number $\tilde \eta$ is related to the reflection $\hat \Pi$ in the plane containing the momentum direction $\hat \Pi|\vec p_z;J^P,\lambda\rangle=\tilde \eta |\vec p_z;J^P,-\lambda\rangle$. It 
   is a good quantum number only for the states with helicity $\lambda=0$ since this transformation inverts helicity; in this case the eigenvalue is $\tilde \eta=P(-1)^J$ where $P$ and $J$ refer to the  rest frame \cite{Thomas:2011rh}. 
  
  Table \ref{tab:irreps} lists irreducible representations $\Lambda$ of the symmetry group $\mathrm{Dic}_4$ relevant for $\vec P\propto \vec e_z$ and corresponding values of helicity $|\lambda|$ and   $\tilde \eta=P(-1)^J$ for the states that this irrep contains  \cite{Thomas:2011rh}. This Table and   $\lambda=-J,..,J$ indicate which $J^P$ appear in a given irrep ($P$ again referring to rest frame):
  \begin{itemize}
  \item $B_1$ and $B_2$ contain $|\lambda|=2$: therefore $J=0,1$ can not contribute, while $J^P=2^\pm,~3^\pm$ contribute
  \item $E_2$  contain $|\lambda|=1$: $J=0$ can not contribute, but all states with $J^P=1^\pm,~2^\pm,~3^\pm$ contribute; this is an example   where particularly many $J^P$ states clutter a single irrep and is therefore challenging
  \item $A_1$ contains $|\lambda|=0$ and $\tilde \eta=P(-1)^J=1$: therefore states with $J^P=0^+,1^-,2^+,3^-$ contribute
  \item $A_2$ contains $|\lambda|=0$ and $\tilde \eta=P(-1)^J=-1$: therefore states with $J^P=0^-,1^+,2^-,3^+$ contribute
  \end{itemize}
  This implies that many states with different $J^P$ appear in a given
  irrep. The eigenstates of light mesons with $|\vec P|=\tfrac{2\pi}{N_L}$
  indeed appear according to this pattern in Figure 8 of
  \cite{Thomas:2011rh}. Charmonium spectrum is denser than the light meson
  spectrum, which implies an even denser spectrum of eigenenergies for each
  irrep in the charmonium case. 
  
 The $C$-parity is a good quantum number also in a moving frame and is a very useful handle to discriminate at least the  states with $C=+1$ and $C=-1$. Consider, for example, the scalar $J^{PC}=0^{++}$ and vector $J^{PC}=1^{--}$ states, that are of significant phenomenological interest:  both appear in  $A_1$, but since they have different $C$-parity, the scalar appears in $\Lambda^C=A_1^{C=+1}$ and the vector appears in a distinct representation $A_1^{C=-1}$.  Note that  $A_1^{C=+1}$ is the only representation that allows to study the scalars with $\vec P\propto \vec e_z$. The challenge is that  the $J^{PC}=0^{++}$ state is inevitably accompanied by  $J^{PC}=2^{++},~1^{-+}$ and $3^{-+}$ states in this representation. 
 
    \section{Operators for spin-identification in moving frames} 

    We identify $J^P$ of the eigenstates by employing the  operators $O^{[J,P,|\lambda|]}_{\Lambda,\mu}(\vec P)$ that are subduced from the continuum operators $O^{J,P,|\lambda|}(\vec P)$ to the irreducible representation $\Lambda$ and row $\mu$, as proposed in \cite{Thomas:2011rh}. Such operators were  found to carry the  memory of the continuum spin-parity $J^P$; they were found to have larger overlap $\langle O|n\rangle$  to the eigenstates $|n\rangle$ with that $J^P$  than to other eigenstates \cite{Thomas:2011rh}.  Comparing magnitudes of  various $\langle O^{[J,P,|\lambda|]}_{\Lambda,\mu}|n\rangle$ therefore helps with identifying $J^P$ of given eigenstates $|n\rangle$. 
    
    The operators, that are particularly useful  for  spin-identification,  are build in three steps \cite{Thomas:2011rh}. The first step is  building  $O^{J,P,M}(\vec P=\vec 0)$  that have good $J^P$ and $M$ at rest and adapting it to the general momentum $\vec P$  
     \begin{align}
O_D^{J,P,M}(\vec P)&=\sum _{m_1,m_2} \!\!\!  CG(1,m_1;1,m_2|J,M) \sum_{\vec x} e^{i\vec P\vec x} ~\bar c (\vec x,t) \Gamma_{m_1} \overleftrightarrow D_{m_2}      c(\vec x,t)~,\nonumber\\
 O_{DD}^{J,P,M}(\vec P)&=\sum _{m_1,m_2,m_3,m_D} \!\!\! \!\!\!\!\! CG(1,m_3;J_D,m_D|J,M) ~CG(1,m_1;1,m_2|J_D,m_D)\sum_{\vec x} e^{i\vec P\vec x} ~\bar c (\vec x,t) \Gamma_{m_3} \overleftrightarrow D_{m_1}   \overleftrightarrow D_{m_2}    c(\vec x,t)~,\nonumber \label{noncorr}
\end{align}

where we show the operators with one and two derivatives, which are
constructed to also have the desired $C$-parity. 
The operators with good helicity $\lambda$ are obtained by the rotation ($R$) of   $\vec e_z$ to $\vec P/|\vec P|$  using the Wigner-matrices ${\cal D}$
\begin{equation}
O^{J,P,\lambda}(\vec P) = \sum_{M}  {\cal D}^{(J)*}_{M,\lambda}(R)~ O^{J,P,M}(\vec P)  ~,\quad \qquad O^{J,P,\lambda}(\vec P=\vec e_z) =  O^{J,P,M=\lambda}(\vec P=\vec e_z)~,
\end{equation}
which is a trivial step for the momentum $\vec P\propto \vec e_z$ employed below. The last step is subducing these operators to the irreducible representation $\Lambda$ and row $\mu$ 
\begin{equation}
O^{[J,P,|\lambda|]}_{\Lambda,\mu}=\sum_{\hat \lambda=\pm |\lambda|} S^{\tilde \eta,\hat \lambda} _{\Lambda,\mu} O^{J,P,\hat \lambda}(\vec P) ~,\qquad  O^{[J,P=+1,|\lambda|]}_{A_1}(\vec P\propto \vec e_z)=O^{J,P=+1,\lambda=M=0}(\vec P\propto \vec e_z)\ \mathrm{for}\ {J=0,2} ~.
\end{equation}
using the subduction matrices $S$ in Table II of \cite{Thomas:2011rh} (we have
verified those $S$ for the symmetry groups $\mathrm{Dic}_4$ and
$\mathrm{Dic}_2$). Below we show the results for the subduction from
$J^P=0^+,~2^+$  to the one-dimensional representation $A_1$, where $S=1$  is trivial. 

\section{Lattice simulation}

In this preliminary study we employ 130 configurations of CLS ensemble (U101)
with   $N_f=2+1$, $V=24^3 \times 128$, $a \simeq 0.0854~$fm and $m_\pi \simeq
280~$MeV. The action includes non-perturbatively O(a)-improved Wilson fermions
with tree-level Symanzik improved gauge action. Fermions and gluons have
periodic boundary conditions in space and open boundary conditions in time.
The correlators are evaluated using the full distillation method
\cite{Peardon:2009gh} with 90 Laplacian eigenvectors.      The charm quarks
are based on $\kappa_c=0.12522$, the gauge links are not smeared and a single source time-slice is used in this study. 

\section{Spectrum and overlaps for $\vec P=\tfrac{2\pi}{L}\vec e_z$, irrep $A_1$ and $C=+1$}

 \begin{figure}[tb] 
  \centering
  \includegraphics[width=6cm,clip]{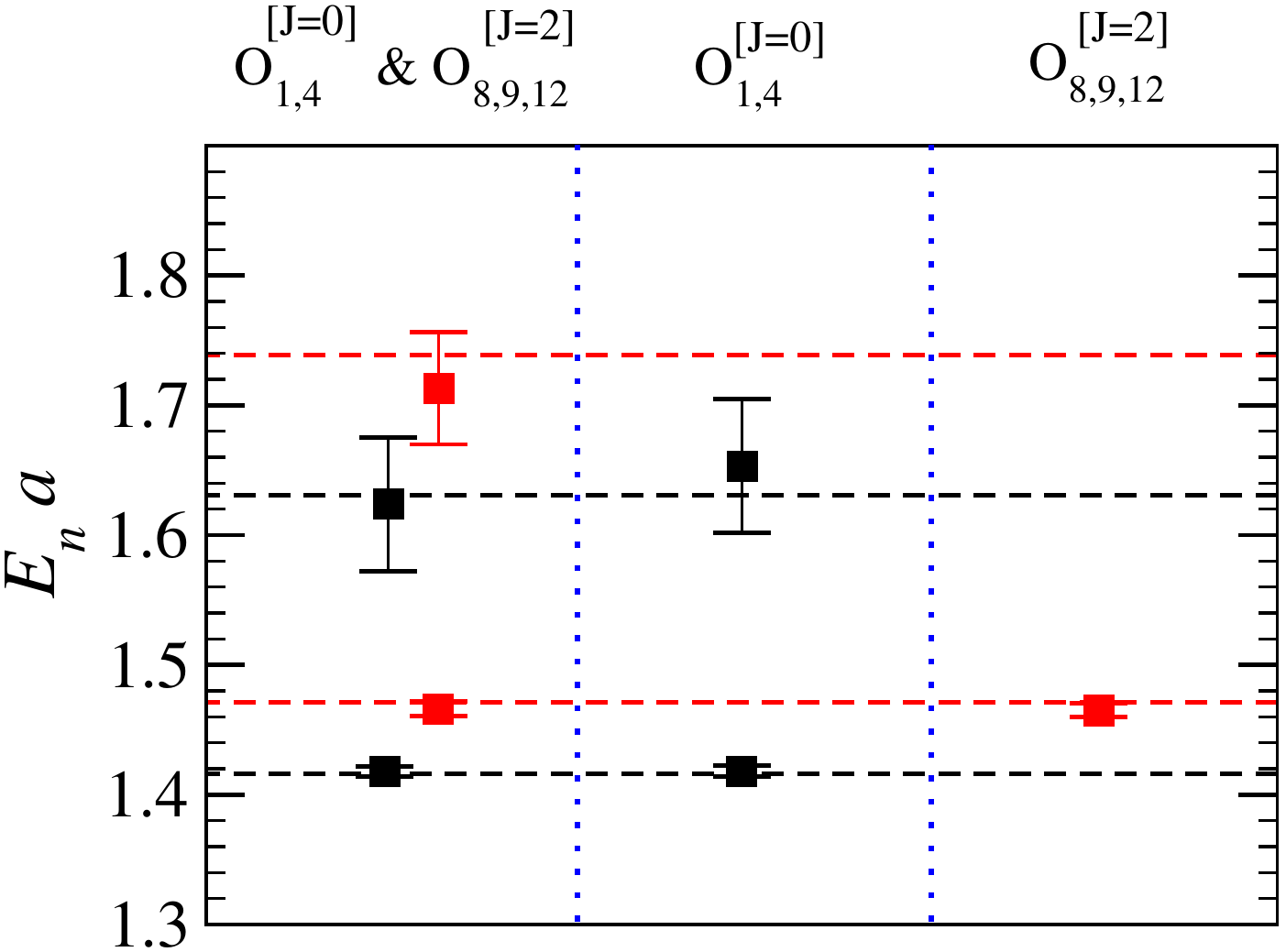}
  \caption{ Eigen-energies $E_{n=1,..}$ for the total momentum $\vec P=\tfrac{2\pi}{N_L}\vec e_z$ in the irreducible representation $A_1^{C=+1}$. Left pane: eigen-energies  resulting from the basis which includes $O^{[J=0]}_{A_1}$ and  $O^{[J=2]}_{A_1}$ operators. Middle pane: energies  resulting from   $O^{[J=0]}_{A_1}$ only. Right pane: energies  resulting from   $O^{[J=2]}_{A_1}$ only. The number of employed operators is given above each pane.  Black and red symbols indicate the eigenstates  identified as $J^P=0^+$ and $J^P=2^+$, respectively, via the employed  spin-identification method. The black dashed lines show $[m_{\chi_{c0}}^2+(2\pi/N_L)^2]^{1/2}$, where $m_{\chi_{c0}}$ correspond to two lowest energies obtained from $\Lambda^P=A_1^+$ with $\vec P=0$. The red dashed lines indicate $[m_{\chi_{c2}}^2+(2\pi/N_L)^2]^{1/2}$, where $m_{\chi_{c2}}$ correspond to two lowest energies obtained from $\Lambda^P=T_2^+$ with $\vec P=0$. }
  \label{fig:spec}
\end{figure}

 \begin{figure}[htb] 
  \centering
 $\qquad \qquad$ \includegraphics[width=9cm,clip]{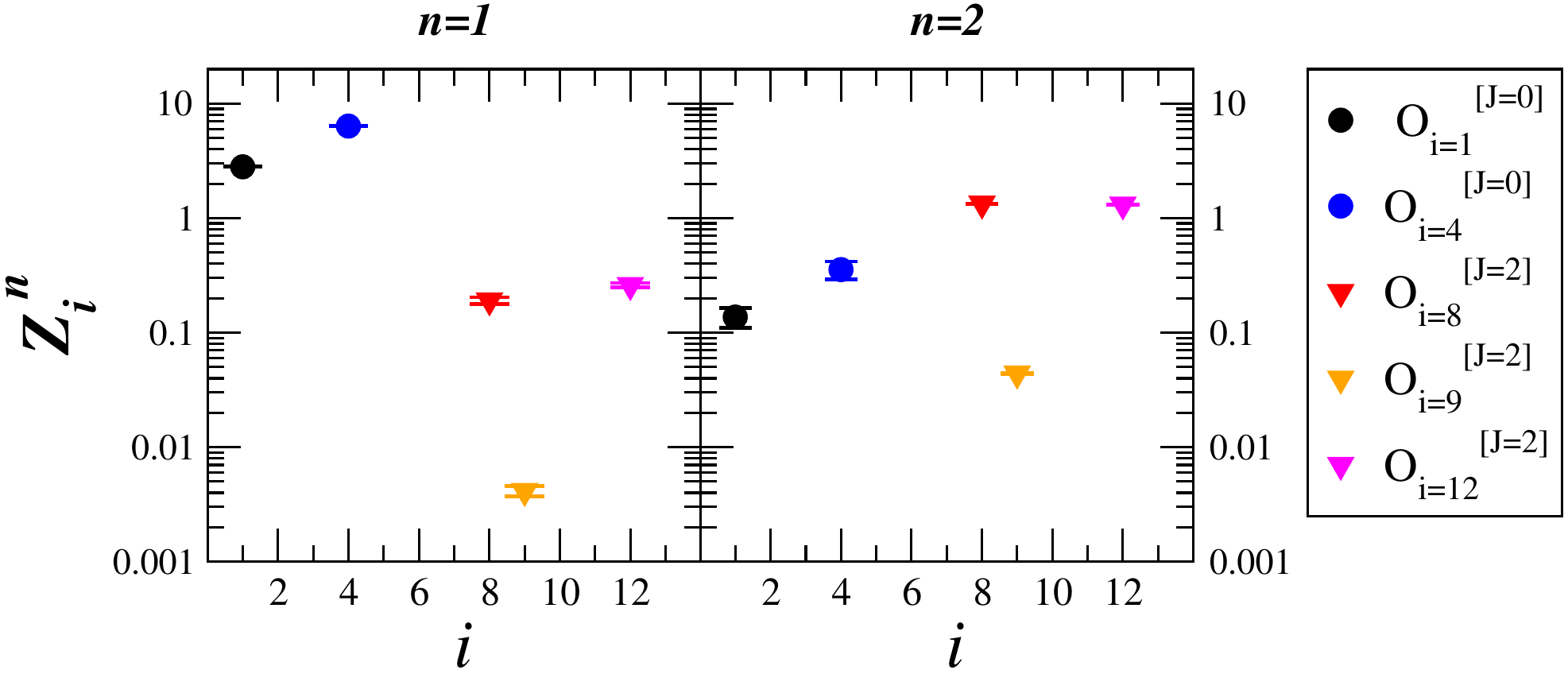}\\
 $\qquad$\\
   \includegraphics[width=7.5cm,clip]{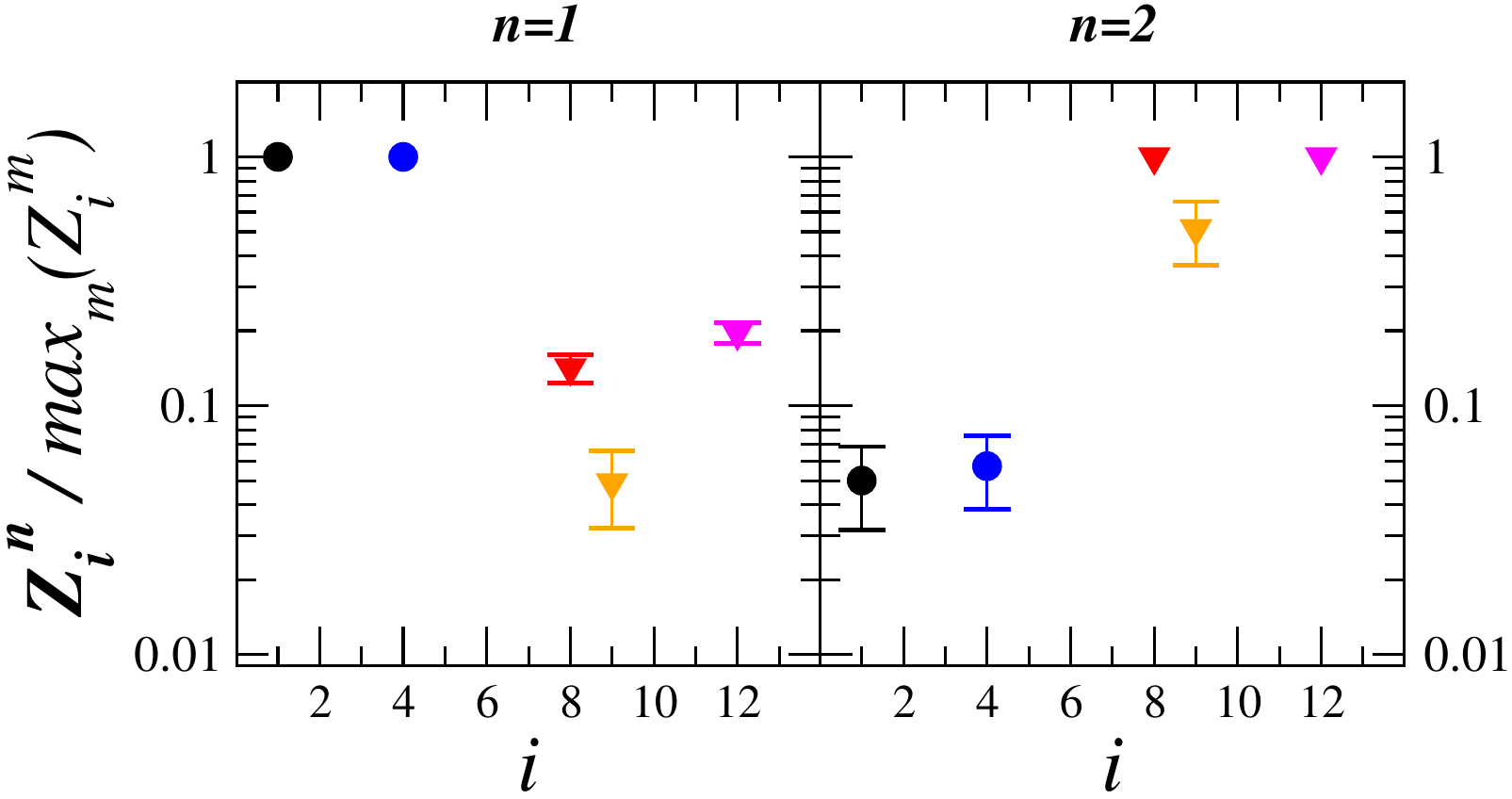}
  \caption{ The overlap factors $Z_i^n=\langle O_i|n\rangle$  for lowest two
    eigenstates $n=1,2$ in the left pane of Figure \ref{fig:spec} resulting
    from the $5\times 5$ correlation matrix based on operators $O_{1,4}^{[J=0]} \&  O_{8,9,12}^{[J=2]}$. The lower figure shows  the normalized overlaps $Z_i^n/max_m Z_i^m$, which are independent of normalization, and are by construction equal to $1$ for the eigenstate ($m$) to which they overlap best.  }
  \label{fig:overlaps}
\end{figure}

As an example, we show the results for $\vec P=\tfrac{2\pi}{L}\vec e_z$, irreducible representation  $A_1$ and positive $C-$ parity. States with $J^{PC}=0^{++},2^{++},1^{-+}$ and $3^{-+}$ can appear as eigenstates in this irrep, as explained in Section \ref{sec:irreps}.  We focus on the states with masses  below $4~$GeV.  The charmonia with $J^{PC}= 1^{-+},~3^{-+}$ are expected to have masses above $4.2~$GeV  (see Figure 3 of  \cite{Cheung:2016bym}), so these quntum numbers are not considered. 
 
The eigenstates $|n\rangle$ with $J^{PC}=0^{++}$ and $2^{++}$ both appear in the same representation $\Lambda^C=A_1^{C=+1}$ and our purpose is to distinguish them based on overlaps  $\langle O^{[J=0]}_{A_1}|n\rangle$ and 
   $\langle O^{[J=2]}_{A_1}|n\rangle$. We employ seven operators $O^{[J=0]}_{i=1,..,7}\equiv  O^{[J=0,P=+,\lambda=0]}_{A_1}$ subduced from continuum $J=0$ and five operators $O^{[J=2]}_{i=8,..,12} \equiv O^{[J=2,P=+,\lambda=0]}_{A_1}$ subduced from continuum $J=2$.

   The left pane of Figure \ref{fig:spec} shows eigen-energies obtained from the GEVP where both $O^{[J=0]}_{A_1}$ and  $O^{[J=2]}_{A_1}$ are used to compute $5\times 5$ correlation matrix\footnote{Five operators are chosen out of twelve, which lead to reasonable plateaus for the lowest four eigenstates.}: four eigenstates are found and at this point it is not clear which are $J=0$ states and which are $J=2$. 
   The identification of $J$   is done based on the overlaps $Z_i^n=\langle O^i|n\rangle$ in  Figure \ref{fig:overlaps}. It shows that a given operator $O_i^{[J=0]}$ has larger overlap $Z_i^n$ with  the state $n=1$ than with $n=2$ (\footnote{This statement is independent of the overall normalization for operator $O_i$ as it compares $Z_i^n$ for various $n$ at the same $i$.}), so eigenstate $n=1$  is identified as $\chi_{c0}$  with $J=0$. Any given operator $O_i^{[J=2]}$ has larger overlap $Z_i^n$ with  the state $n=2$ than with $n=1$, so $n=2$ state is identified as $\chi_{c2}$ with $J=2$. Similarly, $n=3$ and $4$ states are identified as excited $J=0$ and $J=2$ charmonia, respectively, although the  statistical errors on energies and overlaps  are rather large for the current preliminary results.

        This identification of four eigenstates is consistent also with the expectation based on $E=[m_{\chi}^2+(2\pi/N_L)^2]^{1/2}$, which is shown by dashed lines. Here, the masses of $\chi_{c0}$ and $\chi_{c2}$ are extracted from two lowest $\vec P=\vec 0$ states in $\Lambda^P=A_1^+$ and $T_2^+$, respectively. 
    
  For completeness, Figure \ref{fig:spec} shows also eigenstates obtained employing only $O^{[J=0]}_{A_1}$ operators (middle pane) and only $O^{[J=2]}_{A_1}$ operators (right pane).  Given the current statistics and the employed basis, it seems that $O^{[J=0]}_{A_1}$ renders only lowest two $\chi_{c0}$ states, while $O^{[J=2]}_{A_1}$ renders reliably only the ground state $\chi_{c2}$. However, we point out that the middle and right panes do not present a reliable method for identification of $J$. Namely, states with all allowed $J^\prime$ can appear in a given irrep $\Lambda$, although only operators  $O^{[J]}_{\Lambda}$ for specific $J$ are used.  

\section{Conclusions}

The purpose was to identify $J^P$ of the observed charmonium eigenstates in
moving frames. The single-hadron approach was followed and only interpolators
$O=\bar cc$ were used. We employed the spin-identification method proposed in \cite{Dudek:2009qf,Thomas:2011rh}. It   successfully identified the spin-parities $J^P=0^+$ and $J^P=2^+$ of the nearby charmonium states in the irreducible representation $A_1$ for momentum $\vec P=\tfrac{2\pi}{N_L}\vec e_z$. It also worked well for other examples we considered. 
 This will be helpful to spin-identify the narrow states when this study is extended to two-hadron 
scattering, which is in progress on the same ensembles \cite{PiemonteWeishaeupl:2017lat}. 

\vspace{1cm}

\noindent{\bf Acknowledgements}

 The Regensburg group was supported by the Deutsche Forschungs-
gemeinschaft Grant No. SFB/TRR 55. S. P. was supported by Slovenian Research
Agency ARRS (research core funding No. P1- 0035). M. P. acknowledges support from EU under grant no. MSCA-IF-EF-ST-744659 (XQCDBaryons). We thank our colleagues in CLS for the joint effort in the generation of the gauge field ensembles 
which form a basis for the here described computation. The simulations were performed on the Regensburg iDataCool cluster
and the SFB/TRR 55 QPACE2 \cite{Arts:2015jia} and QPACE3 machines. Part of the simulations 
were preformed at the local cluster at the Departument of Teoretical physics at Jozef Stefan Institute. 


\begin{thebibliography}{7}

\bibitem{Lang:2015sba}
C.B. Lang, L.~Leskovec, D.~Mohler, S.~Prelovsek, JHEP \textbf{09}, 089 (2015),
  \texttt{1503.05363}

\bibitem{PiemonteWeishaeupl:2017lat}
S.~Piemonte, S.~Weishaeupl, G.~Bali, S.~Collins, D.~Mohler, M.~Padmanath,
  S.~Prelovsek, \emph{{Charmonium resonances on the lattice}}, in
  \emph{Proceedings, \href{http://inspirehep.net/record/1425631}{35th
  International Symposium on Lattice Field Theory (Lattice2017)}: Granada,
  Spain}, indico IDs 138 and 183, to appear in EPJ Web Conf.

\bibitem{Cheung:2016bym}
G.K.C. Cheung, C.~O'Hara, G.~Moir, M.~Peardon, S.M. Ryan, C.E. Thomas, D.~Tims
  (Hadron Spectrum), JHEP \textbf{12}, 089 (2016), \texttt{1610.01073}

\bibitem{Dudek:2009qf}
J.J. Dudek, R.G. Edwards, M.J. Peardon, D.G. Richards, C.E. Thomas, Phys. Rev.
  Lett. \textbf{103}, 262001 (2009), \texttt{0909.0200}

\bibitem{Thomas:2011rh}
C.E. Thomas, R.G. Edwards, J.J. Dudek, Phys. Rev. \textbf{D85}, 014507 (2012),
  \texttt{1107.1930}

\bibitem{Peardon:2009gh}
M.~Peardon, J.~Bulava, J.~Foley, C.~Morningstar, J.~Dudek, R.G. Edwards,
  B.~Joo, H.W. Lin, D.G. Richards, K.J. Juge (Hadron Spectrum), Phys. Rev.
  \textbf{D80}, 054506 (2009), \texttt{0905.2160}

\bibitem{Arts:2015jia}
P.~Arts et~al., PoS \textbf{LATTICE2014}, 021 (2015), \texttt{1502.04025}

\end{thebibliography}

\end{document}